\pdfoutput=1

\documentclass[11pt,dvipsnames]{article}

\usepackage[preprint]{acl}

\usepackage{times}
\usepackage{latexsym}

\usepackage[T1]{fontenc}

\usepackage[utf8]{inputenc}

\usepackage{microtype}

\usepackage{inconsolata}

\usepackage{graphicx}

\usepackage{amsmath,amssymb}
\usepackage{multirow}

\usepackage{booktabs}

\def\*#1{\mathbf{#1}}

\title{A Text-Based Recommender System that Leverages\\Explicit Affective State Preferences}

\author{Tonmoy Hasan \and Razvan Bunescu\\
         Department of Computer Science\\
         University of North Carolina at Charlotte\\
         \texttt{thasan1,rbunescu@charlotte.edu}
         }

\begin{document}
\maketitle
\begin{abstract}
The affective attitude of liking a recommended item reflects just one category in a wide spectrum of affective phenomena that also includes emotions such as entranced or intrigued, moods such as cheerful or buoyant, as well as more fine-grained affective states, such as "pleasantly surprised by the conclusion". In this paper, we introduce a novel recommendation task that can leverage a virtually unbounded range of affective states sought explicitly by the user in order to identify items that, upon consumption, are likely to induce those affective states.  Correspondingly, we create a large dataset of user preferences containing expressions of fine-grained affective states that are mined from book reviews, and propose a Transformer-based architecture that leverages such affective expressions as input. We then use the resulting dataset of affective states preferences, together with the linked users and their histories of book readings, ratings, and reviews, to train and evaluate multiple recommendation models on the task of matching recommended items with affective preferences. Experiments show that the best results are obtained by models that can utilize textual descriptions of items and user affective preferences.
\end{abstract}

\section{Introduction and Motivation}

Traditional recommender systems (RS) leverage user preferences that are implicit in user-item rating histories in order to personalize rankings of the items presented to the user \cite{resnick:acm97,park:esa12}. As long as sufficient user-item data is available, this recommendation approach is convenient for the user as it requires little to no interaction other than providing rating feedback. This minimal interaction approach, however, can be slow to track changes in user preferences and imprecise for users with diverse preferences. Conversational recommender systems (CRS) \cite{thompson_personalized_2004} can elicit more precise preferences by engaging in a natural dialogue with the user, an approach that more recently takes advantage of fine-tuned or appropriately prompted LLMs \cite{zhao_recommender_2024,lian_recai_2024}. Besides their ability to carry goal driven conversations, large scale pre-trained language models also bring the advantage of having an extensive implicit memory of items and their descriptions \cite{penha_what_2020}, which benefits content-based recommendations, as well as knowledge of correlations between item consumption histories and user preferences, which in theory could provide useful collaborative-filtering signals \cite{he_large_2023}. To support the training and evaluation of CRS models, a number of conversational recommendation datasets have been created using either crowd-sourcing or scrapping of real interactions on dedicated internet forums. In the movie domain, for example, datasets such as INSPIRED \cite{hayati_inspired_2020} and ReDIAL \cite{li_towards_2018} use a crowd-sourcing approach where workers play the roles of seekers and recommenders, whereas the Reddit-Movie \cite{he_large_2023} dataset contains naturally occurring dialogues where Reddit users seek and offer recommendations in the real world. Table~\ref{tab:epistemic} shows in narrative form prototypical examples of user preferences that were elicited in conversational recommendation dialogues. These examples illustrate a predominant aspect of CRS datasets: with very few exceptions, the elicited preferences refer to objective features of an item (colored in \textcolor{teal}{\it teal}) or to items the user has consumed in the past. For example, users express a like or dislike attitude towards movie genres (comedies, fantasy, sci-fi, or thriller), movie directors (Kubrik), source material (DC Comics), plot and content (romantic vibe or inventive). We call these item features {\it objective} as they are largely independent of the user, e.g., given an arbitrary movie, most users would agree on whether the movie is a thriller, and similarly the identity of the movie director should be uncontroversial. The {\it subjective} aspect of the elicited preferences is then due solely to the user's attitude of liking or disliking particular items or their objective features.
\begin{table}[t]
    \centering \small
    \begin{tabular}{p{\columnwidth}}
    \hline
       $\blacktriangleright$ I love \textcolor{teal}{\it superhero} movies. Last night I saw Shazam and I loved the \textcolor{teal}{\it comedy} part of it too. I usually don't like \textcolor{teal}{\it DC movies} because they lack \textcolor{teal}{\it humor}, but this movie had that. \\
       $\blacktriangleright$ Can you recommend a movie like A Clockwork Orange? I liked that it was a \textcolor{teal}{\it Kubrick movie} and that it was inventive. I didn't like that \textcolor{teal}{\it the "future" portrayed was a little dated}. \\
       \hline
       $\blacktriangleright$ I'm into \textcolor{teal}{\it Fantasy}. Especially \textcolor{teal}{\it high fantasy}. I've seen Highlander. I've seen them all actually. I enjoyed them. The Matrix was great! But I didn't care for The Matrix Revolutions. It \textcolor{teal}{\it drifted from the original story} in my opinion. Blade Runner is definitely my kind of movie. It's the perfect combo of \textcolor{teal}{\it scifi and thriller}.\\
       \hline
       $\blacktriangleright$ Any movies like before Sunrise trilogy? I'm looking for movies where \textcolor{teal}{\it a couple or strangers explores the ideas and world around them} and also has \textcolor{teal}{\it a romantic vibe} to it. \\
       $\blacktriangleright$ I've always enjoyed movies that have \textcolor{teal}{\it dark themes} - movies that \textcolor{violet}{\it make you uncomfortable/squirm} be it from a psychological standpoint or from just sheer brutality. Obviously with a \textcolor{violet}{\it compelling story} or \textcolor{Violet}{\it thought provoking ideas}.\\
       \hline
    \end{tabular}
    \caption{Examples of preferences from INSPIRED (top), ReDIAL (middle), and Reddit-Movie (bottom). Objective features are shown colored in \textcolor{teal}{\it teal}.}
    \label{tab:epistemic}
\end{table}

\begin{table}[t]
    \centering
    \small
    \begin{tabular}{p{\columnwidth}}
        \hline
        $\blacktriangleright$ I want a book that \textcolor{violet}{\it makes me smile at times}, but that also \textcolor{violet}{\it brakes my heart}. A book where the main character is a child narrator. The book will \textcolor{Violet}{\it make me think about it for a long time}.\\
        $\blacktriangleright$ I am looking for a young adult horror book that doesn't hold back. It is 100\% shocking, scary, and over the top. I would like to \textcolor{violet}{\it feel thrilled, disturbed, and completely entranced} while reading. The book has sympathetic characters in horrible but believable situations. \textcolor{violet}{\it I will not want to put the book down except to pace around the room to shake off the chills.}\\
        $\blacktriangleright$ I am looking for a book \textcolor{Violet}{\it unlike any I've ever experienced before}. The book's \textcolor{violet}{\it emotional impact is so heavy that, even though I'm super into it, I will have to stop for a couple of days before finishing it.}\\
        $\blacktriangleright$ I am drawn to books about complex social issues especially those that impact families. Characters that are real, who are developed well enough so that \textcolor{violet}{\it I can feel their emotions} and a strong story line make a book exceptional for me. 
        It is a book that \textcolor{violet}{\it I will not want to put down} and which evokes \textcolor{violet}{\it a depth of emotions that takes me by surprise.}\\
        \hline 
    \end{tabular}
    \caption{Examples of affective preferences generated from Goodreads reviews. Expressions of affective states are colored in \textcolor{violet}{\it violet}, affective-cognitive states in \textcolor{Violet}{\it blue}.}
    \label{tab:affective}
\end{table}
The attitude of liking an item is the only type of affective state that is explicitly targeted by sentiment analysis or recommender systems. However, following Scherer's typology of affective states \cite{scherer_what_2005}, there are not one, but three major types of affective states that an item can impress on a user: {\it attitudes}, {\it moods}, and {\it emotions}. Attitudes are relatively enduring, affectively colored beliefs and dispositions towards items. The affective states induced by a salient attitude are generally weak in intensity and can be labeled with terms such as liking, disliking, loving, hating, valuing, or desiring. Moods are diffuse, low intensity affect states, characterized by a relative enduring predominance of certain types of subjective feelings. Examples are being cheerful, gloomy, listless, depressed, or buoyant. Emotions are comparatively shorter in duration but higher in intensity, and can have a strong behavioral impact, often altering ongoing behavior sequences and triggering the generation of new goals and plans. Examples include {\it utilitarian} emotions, such as anger, sadness, joy, fear, pride, guilt, or shame, and {\it aesthetic} emotions, such as wonder, awe, admiration, fascination, bliss, ecstasy, harmony, rapture, or solemnity. They can generally be elicited by a wide array of both external and internal stimulus events, such as natural phenomena, the behavior of others or one's own behavior, and sudden neuroendocrine or physiological changes. Especially relevant to this work, emotions can also be elicited by memories or images that might come to one's mind, such as the ones evoked by reading a book. These imagined representations of events are often sufficient on their own to generate strong emotions \cite{goldie_life_2004}. Table~\ref{tab:affective} shows examples of {\it affective} states (colored in \textcolor{violet}{\it violet}) that were induced by reading a book, sampled from reviews in the Goodreads dataset \cite{wan2018goodreads}. Also shown are examples of more complex {\it cognitive} states (colored in \textcolor{Violet}{\it blue}), named as such because they can be seen as mixing a strong affective component with other cognitive aspects, such as reasoning, attention, memory, or decision making. Note that affective states are much scarcer in Table~\ref{tab:epistemic}, showing up only in a sample from  Reddit-Movie. Compared to the other two datasets, which were created from people pretending to seek movie recommendations, Reddit-Movie contains real world, genuine recommendation requests. As such, we believe that recommender systems would provide significantly higher user satisfaction if they were able to {\it elicit} and {\it match} a wider range of desired affective states. Correspondingly, the main contribution of this paper is a novel recommender system framework where users can express a wide range of affective state preferences.

\begin{figure*}[t]
    \centering
    \includegraphics[width=1.0\textwidth]{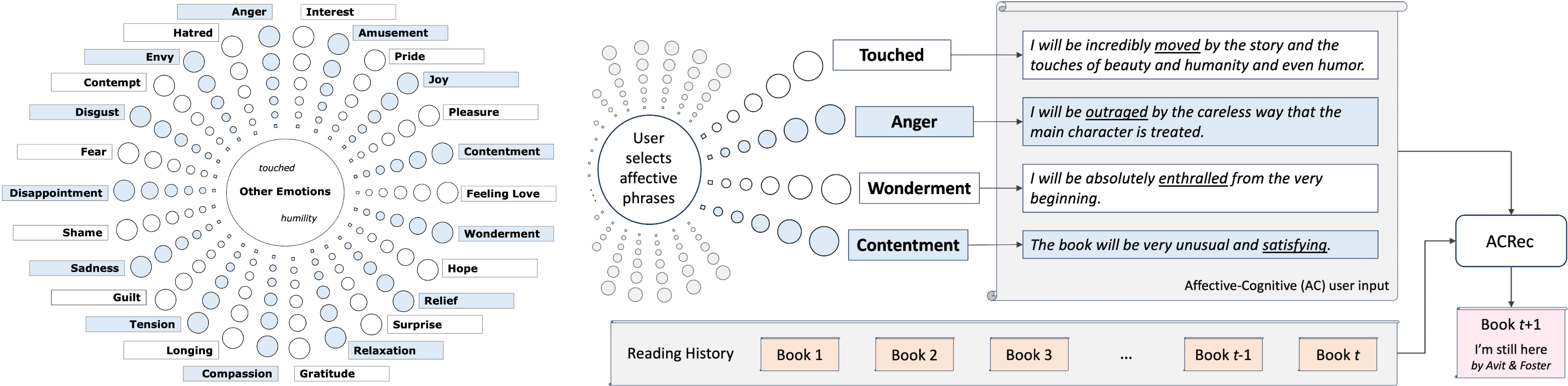}
    \caption{Selection of AC statements, starting from top emotion categories organized around the Emotion Wheel. Left: The proposed emotion wheel adapted from the Geneva Emotion Wheel. Right: The user selects emotion categories, where for each category they have the option of selecting fine-grained AC statements.}
    \label{fig:wheel-acrec}
\end{figure*}

The structure of the paper is as follows: in Section~\ref{sec:acrec-task} we introduce the task of affective-cognitive recommendations; in Section~\ref{sec:acrec-framework} we present the architecture of the corresponding recommender system; in Section~\ref{sec:dataset} we introduce the book recommendation dataset used for training and evaluations, and the procedure used for generating affective-cognitive statements from reviews; in Section~\ref{sec:evaluation} we present experimental evaluations of the proposed architecture and a discussion of the results; in Section~\ref{sec:related} we discuss related work. The paper ends with thoughts on limitations and conclusion.

\section{The Affective-Cognitive Recommender}
\label{sec:acrec-task}

An Affective-Cognitive Recommender (ACRec) system takes as input data about a user, such as their history of readings, ratings, and reviews, as well as a description of the affective-cognitive (AC) states that they seek to experience upon consuming an item. Based on this input, ACRec computes a ranked list of items, where items at the top are expected to induce the AC states desired by the user, as expressed in the AC description part of the input. To input the AC description, the user can use the two modes below, separately or in combination:
\begin{enumerate}
    \item {\bf Generation}: the user generates the AC description as free form text (Table~\ref{tab:affective}).
    \item {\bf Selection}: the user selects one or more expressions of affective states from a predefined repository of AC statements (Figure~\ref{fig:wheel-acrec}).
\end{enumerate}
The four AC descriptions shown in Table~\ref{tab:affective} belong to the Generation mode. While useful for specifying objective features, as shown in Table~\ref{tab:epistemic}, for most users the generation mode can be much less conducive to eliciting expressions of affective states, due to the difficulty of  expressing fine-grained emotions that one would like to experience from a book they have yet to read. As such, we also propose a Selection mode that enables the user to select expressions of desired affective states from a predefined, rich repository of AC statements. To this end, we have created a large dataset of expressions of fine-grained affective states that are mined from book reviews, as detailed in Section~\ref{sec:dataset}. Given the thousands of AC statements contained in this dataset, for the selection process to be effective it is important that the AC statements are organized in an ontology that users can navigate easily. As such, we propose that the repository of AC statements is organized at the top level around a wheel of emotion categories, as shown at the left of Figure~\ref{fig:wheel-acrec}. We created this Emotion Wheel by augmenting the Geneva Emotion Wheel (GEW) \cite{gew:12} with 6 additional emotion categories that were originally mentioned in \cite{scherer_what_2005} and that were observed to be expressed in user reviews: Contentment, Gratitude, Hatred, Hope, Relaxation (Serenity), and Tension (Stress). GEW is a theoretically derived and empirically tested instrument that was designed to measure emotional reactions to objects, events, and situations. The original 20 discrete emotion categories (or families) are organized around a circle, where the horizontal dimension indicates the emotion valence (from negative to the left to positive to the right) and the vertical dimension indicates control (from low at the bottom to high at the top). Additionally, spokes in the wheel 
correspond to different levels of intensity for each emotion family from low intensity (towards the center) to high intensity (toward the circumference), where each intensity level can be associated with typical words that are used to express emotions at that level. The center of the wheel is a catch all category for emotions that do not belong to any of the main categories.

Figure~\ref{fig:wheel-acrec} illustrates ACRec in the Selection mode, where the user starts by selecting 4 emotions categories. For each category they have the option of selecting an emotion word corresponding to one of the intensity levels on the spoke. The system would then present them with the set of AC statements in the repository that correspond to that emotion category (and word if selected), possibly subcategorized according to the source of emotion, e.g. characters, writing style, or events in the book. The AC statements selected by the user using the process above would form the AC description. 

Together with the user history of readings, reviews, and ratings, the AC description would then be provided as input to the ACRec recommendation model, which is trained to compute a ranking of all the books with respect to how likely they are to trigger in the user the affective states contained in the AC description. In the following section we introduce the architecture of an AC recommendation model that is specifically designed to leverage AC descriptions elicited from users.



\begin{table}[b]
\centering \small
\begin{tabular}{ c | p{5.5cm} }
 \hline
Notations & Explanations \\
 \hline
 $\mathcal{U}, \mathcal{B}$ &  User and item set \\ 
 $\mathcal{B}^u$ & Set of items that user $u$ has read, rated and written reviews for\\
 ${\mathbf{d}^u_i} \in \mathbb{R}^d$ & Embedding of $i$-th item's original description\\
  ${\mathbf{e}^u_i} \in \mathbb{R}^d$ & Embedding of $i$-th item's extended description\\
 ${\mathbf{r}^u_i} \in \mathbb{R}^d$ & Review embedding of $i$-th item\\
 ${\mathbf{g}^u_i} \in \mathbb{R}^n$ & Rating embedding of $i$-th item\\
 ${\mathbf{c}^u_i}\in \mathbb{R}^l$ & Concatenated embedding of $i$-th time step\\
 $\mathbf{sp}_t^u \in \mathbb{R}^l$ & User $u$'s short-term preference at time step $t$\\
 $\mathbf{lp}^u_t\in \mathbb{R}^l$ & User $u$'s long-term preference at time step $t$\\
 $y_{b}^u$ & Recommendation score for candidate book $b$ for user $u$\\
\hline
\end{tabular}
\caption{Key notations and their description.}
\label{tab:notations}
\end{table}

\section{The AC Recommendation Model}
\label{sec:acrec-framework}

\begin{figure*}[ht]
    \centering
    \includegraphics[width=\textwidth]{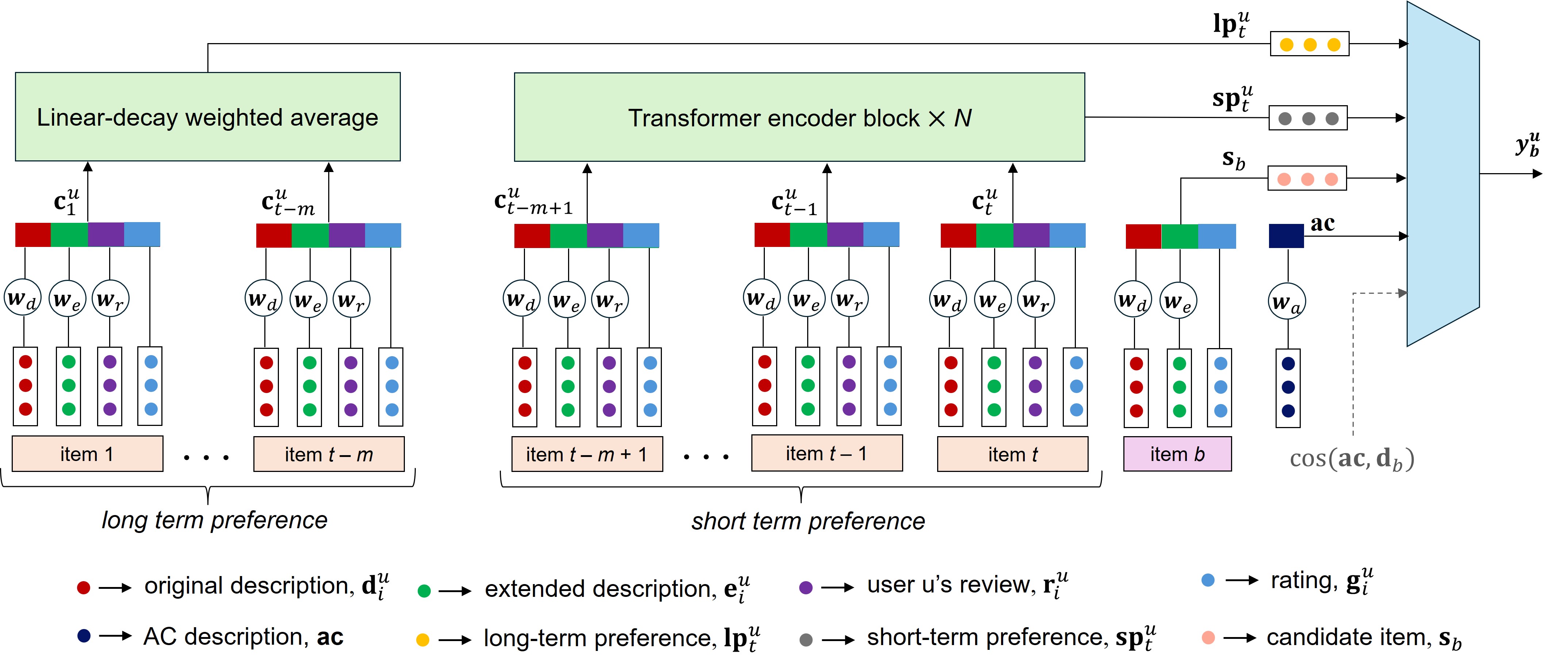}
    \caption{Overall architecture of the proposed ACRec model.}
    \label{fig:framework}
\end{figure*}

Given the inherent subjective nature of affective states, the recommender system needs to have a good model of user preferences and expectations. These will be inferred from the user data, which will consist of the history of consumed items, ratings, and reviews (notation summarized in Table~\ref{tab:notations}). Let $\mathcal{U}$ be the set of users and $\mathcal{B}$ the set of items, alternatively referred to as books. For a user $u \in \mathcal{U}$, let $\mathcal{B}^u = \{b^u_1, b^u_2, \dots, b^u_{t-1}, b^u_t\} \subseteq \mathcal{B}$ be a sequence of items in chronological order that the user $u$ has consumed, rated, and written reviews for, up to time step $t$. Each item in $\mathcal{B}$ is associated with an original description and another review-based extended description. Given data about a user $u$ up to and including the current time step $t$, an AC description $ac$, and an arbitrary book $b$, the task of the ACRec system is to compute a recommendation score $y_{b}^u(ac, b)$ that should reflect how well the book $b$ aligns with the AC description $ac$ provided by the user $u$.

The overall architecture of the proposed ACRec model is shown in Figure \ref{fig:framework}. The recommendation score $y_{b}^u(ac, b)$ is calculated by a fully connected network with one or more hidden layers that takes as input representations of long-term user preferences $\mathbf{lp}_t^u$, short-term user preferences $\mathbf{sp}_t^u$, the AC description $\mathbf{ac}$, and the representation $\mathbf{s}_b$ of an arbitrary book $b$. Additionally, the model can also accommodate as input the cosine similarity $cos(\mathbf{ac}, \mathbf{d}_b)$ between the AC description and the book description.

Each book is associated two textual descriptions: the {\it original} description from the Goodreads dataset, and an {\it extended} description created from the concatenation of user reviews for that book. It is important to note that reviews from training users and testing users are never used as part of extended descriptions. A book that the user $u$ read, reviewed, and rated at time step $t$ is then represented as the concatenation of 4 embeddings $\mathbf{c}^u_t = [\mathbf{d}^u_t;\mathbf{e}^u_t;\mathbf{r}^u_t;\mathbf{g}^u_t]$, where $\mathbf{d}^u_t$ is an embedding of the original book description, $\mathbf{e}^u_t$ is an embedding of the extended book description, $\mathbf{r}^u_t$ is an embedding of the user's review of the book, and $\mathbf{g}^u_t$ is an embedding of the rating (grade) that the user gave to the book. We first use Jina Embeddings v3 \cite{sturua2024jina}, a text embedding model that produces 1024-dimensional embeddings of text containing up to 8192 tokens, with the flexibility to reduce dimensionality without compromising performance. To obtain $\mathbf{d}^u_t$, $\mathbf{e}^u_t$, and $\mathbf{r}^u_t$, we first generate their 768-dimensional Jina embeddings and reduce them to 41-dimensional embeddings using learned projection matrices $W_d$, $W_e$, and $W_r$, respectively. Size 5 embedding are also learned for each of the possible 5 rating values. When concatenated together, the 4 embeddings in $\mathbf{c}^u_t$ add up to a size of $3 \cdot 41 + 5 = 128$, to match the tuned hidden dimension used in the Transformer block. Overall, the user data up to the current time step $t$ can then represented as the sequence $\langle \mathbf{c}^u_1, \mathbf{c}^u_2,..., \mathbf{c}^u_{t-1}, \mathbf{c}^u_t\rangle$.

\subsection{User Preference Modeling}
\label{sec:preferences}

We model short-term and long-term preferences by partitioning the user's time series data in two parts:
\begin{enumerate}
    \item {\bf Short term preferences} $\mathbf{sp}_t^u$ are modeled by running a Transformer over the last $m$ items $\langle\mathbf{c}^u_{t}, \mathbf{c}^u_{t-1}, \dots, \mathbf{c}^u_{t-m+1}\rangle$ in this reversed order.
    \item {\bf Long term preferences} $\mathbf{lp}_t^u$ are created from a linearly weighted average of the remaining $t - m$ items $\langle\mathbf{c}^u_1, \mathbf{c}^u_2, \dots, \mathbf{c}^u_{t-m}\rangle$.
\end{enumerate}
The short-term preference embedding $\mathbf{sp}_t^u$ is set to be the hidden state computed by the last Transformer block for time step $t$. We use a Transformer model with hidden size of 128, 4 blocks, and the standard, fixed positional embeddings \cite{NIPS2017_3f5ee243}.

The long-term preferences $\mathbf{lp}_t^u$ are calculated as:
\begin{equation}
    \mathbf{lp}^u_t = \sum_{k=1}^{t-m} \alpha_k \mathbf{c}^u_k
\label{eq:long_term}
\end{equation}
where the linearly decaying weights are set as $\alpha_k = 2k / ((t - m)(t - m + 1))$. The weights sum up to 1 and ensure that more recent items are assigned larger weights compared to more distant items.

\subsection{Recommendation Score Calculation}
\label{sec:score}

Besides the long-term and short-term preferences of the user at time step $t$, computing the recommendation score for a book $b \in \mathcal{B} - \mathcal{B}_t^u$ that the user has not read yet requires a representation of the book $\mathbf{s}_b$ as well an embedding $\mathbf{ac}$ of the AC description provided by the user. The book representation $\mathbf{s}_b$ is a concatenation of its original description embedding, its extended description embedding, and the embedding of the maximum rating value of 5. These embeddings are calculated using the same procedure that was used for computing the embeddings in the item representations $\mathbf{c}_t^u$. Note that we do not use a book review embedding in $\mathbf{s}_b$, as the candidate book $b$ has not been read by the user. The AC description is embedded into a vector $\mathbf{ac}$ by projecting its Jina embedding using a separate projection matrix $W_a$.

The representations of the 4 inputs, namely the long-term and short-term preferences $\mathbf{lp}_t^u$ and $\mathbf{sp}_t^u$, the candidate book $\mathbf{s}_b$, and the AC description $\mathbf{ac}$, are then passed as input to a fully connected network with two hidden layers and one linear output layer that computes the recommendation score as shown below, where $f$ is the ramp function:
\begin{align} 
y_{b}^u &= \mathbf{w}_3 f(\mathbf{W}_2 f (\mathbf{W}_1[\mathbf{lp}^u_t; \mathbf{sp}_t^u; \mathbf{s}_b; \mathbf{ac}]))
\label{eq:acrec_v2}
\end{align}
We also experimented with one and three hidden layers, however two hidden layers obtained the best performance on validation data.

\subsection{Training Objective}
\label{subsec:model_training}

To train the ACRec model, we employ the Bayesian personalized raking loss \cite{10.5555/1795114.1795167} shown below:
\begin{equation} \label{eq:obj1}
    \mathcal{L(\mathbf{\Theta})} = - \sum_{u\in U} \frac{1}{T_u}\sum_{t=1}^{T_u} \frac{1}{K}\sum_{k = 1}^K\log \sigma (y_{b}^u - y_{k}^u)
\end{equation}
where $T_u$ is the number of training steps, $K$ is the number of negative samples, $y_{b}^u$ is the recommendation score for the positive sample, and $y_{k}^u$ is the recommendation score from a negative sample. As will be explained in Section~\ref{sec:dataset} below, as positive sample we use the book know to match the AC description for the user, whereas negative samples will be drawn at random at each gradient update step from books that the user has not read. The loss is minimized using the Adam Optimizer \cite{kingma2014adam} with a batch size of 1.

\section{The AC Book Recommendation Dataset}
\label{sec:dataset}

To train and evaluate the ACRec model, we create a large dataset that maps recommended books to AC descriptions of user preferences containing expressions of fine-grained affective states mined from book reviews contained in the Goodreads dataset \cite{10.1145/3240323.3240369}.

\subsection{User Reading Histories}

We extract from the Goodreads dataset book reading histories corresponding to a total of 1,000 users. In order to address a diverse range of reading histories, we randomly select 100 users who have read between \([20\text{-}50]\) books. We then randomly select other 100 users for each subsequent range \([51\text{-}100]\), \([101\text{-}150]\), and so on up to \([451\text{-}500]\) books, resulting in a total of 1,000 users. 

Given a user reading history, we identify a time step as {\it useful} if the user's book rating is 4 or 5, the number of tokens of the review written by the user is at least 20, and the combined token count of the original book description and the extended book description is at least 250. 
We skip the first 15 time steps from the user's history, considering them a burn-in set for learning the user's reading preferences. We then select a maximum of 20 useful time steps per user, spread evenly across the user's reading history. These time steps from all 1,000 training users are then split into training, validation, and testing, as described in Section~\ref{sec:train-test}. Overall, the reading histories of the 1,000 users in the dataset contained 249,326 times steps in total, covering 142,892 books and 11,588 useful steps.

\subsection{Affective-Cognitive Statements}
\label{sec:ac-extraction}

We used GPT-4o to extract statements of affective-cognitive (AC) preferences from book reviews, in two phases. In Phase 1, we instruct GPT-4o to process the book reviews into statements that do not contain identifiable information such as book titles, author names, character names, direct quotations, chapter or page references, or sentences expressing negative sentiment. The extracted statements are then categorized into three types: Affective (A), Cognitive (C), and Affective-Cognitive (AC). An Affective statement conveys the reader’s emotional or mood-based response to the book; a Cognitive statement highlights factual, structural, or stylistic aspects, focusing on objective features without reference to emotional reactions; and an Affective-Cognitive statement integrates both, expressing emotional responses tied to specific elements of the book’s content or form. Each statement should be self contained and should read as if written before reading the book, by someone seeking to learn from and experience what is described in the review, without having any prior knowledge of the actual book. A manual evaluation of 100 extracted AC statements, based on the rubrics in Appendix~\ref{sec:dataset_appendix}, showed that 99 adhered to the guidelines, with one exception containing a mildly negative sentiment sentence. However, we also observed that, likely due to the large size of the prompt, which includes both detailed guidelines and in-context examples, combined with the often lengthy nature of the book reviews, GPT-4o occasionally misclassified C phrases as A phrases. To improve the classification accuracy, in Phase 2 we instructed the LLM to fix its initial categorization of AC statements by providing it with instructions and discriminative examples from each category. Following this refinement, we manually evaluated a random sample of 100 extracted statements in terms of how precise the model was in distinguishing phrases with affective content (A + AC) from phrases with only objective content (C). We found that GPT-4o achieved an overall precision of 88\%. Overall, this approach resulted in 12,522 affective (A), 1,252 affective-cognitive (AC), and 45,935 cognitive (C) statements.

Finally, to support the Selection mode described in Section~\ref{sec:acrec-task}, all statements that contain affective content (A + AC) were mapped to the 26 + 1 categories on the Emotion Wheel, using the LLM-based approach described in Appendix~\ref{sec:wheel-classification-appendix}.

Prompt details and in-context examples for all these tasks are provided at the following link\footnote{\url{https://github.com/Ton-moy/ACRec_Taxonomy}}.

\subsubsection{Training, Validation, and Testing}
\label{sec:train-test}

For each of the 1,000 users in the dataset, we use the last (most recent) useful time step from the user's history for testing, while the second-to-last acceptable time step is used for validation. All remaining (earlier) useful time steps are used for training. During training, for each user, we sample $K = 10$ books that the user has not read to use as negative sample in the Bayesian ranking loss computation. For each training, validation, and test time steps, we utilize as input an AC description that aggregates all A, AC, and C statements that were extracted (with the approach described in Section~\ref{sec:ac-extraction}) from the review that the user wrote for the book at that time step, whereas the book itself is used as the ground truth positive sample.

Note that this is a retrospective evaluation, which means that elicitation of AC preferences is not necessary because they are already available in the review that the user wrote for the book recommended at timestep $t+1$. Thus, AC statements are extracted from the review of the book that was read at time $t+1$ and are used as input, together with the reading history up to time $t$, to recommend a book to read at time $t+1$. The book that the user actually read at time $t+1$ (whose review was used to extract AC statements) is used as the ground truth book for evaluation. Naturally, in practice, this retrospective use is not possible because the review of a book does not exist before recommending and reading the book. Therefore, the user will have to specify their AC preferences, e.g., by using the generation or selection modes described in Section~\ref{sec:acrec-task}.


\section{Experimental Evaluation}
\label{sec:evaluation}

\begin{table*}[t]
    \centering
    \small
    \begin{tabular}{|l|l|rrrrr|rrrr|}
    \hline
    Description & Recommender & \multicolumn{5}{c|}{Hit Ratio (HR)} & \multicolumn{4}{c|}{NDCG} \\
    \cline{3-11}
    Type & System & @1 & @5 & @10 & @50 & @100 & @5 & @10 & @50 & @100 \\
    \hline
         & SASRec & 0.1 & 0.7 & 1.1 & 2.4 & 3.2 & 0.4 & 0.5 & 0.8 & 0.9 \\
         & BERT4Rec & 0.1 & 0.1 & 0.2 & 0.8 & 1.4 & 0.1 & 0.1 & 0.2 & 0.3 \\
    A + AC + C  & UniSRec & 0.0 & 0.1 & 0.3 & 1.6 & 3.2 & 0.0 & 0.1 & 0.4 & 0.6 \\
         & RecFormer & 0.2 & 1.8 & 2.8 & 5.7 & 8.8 & 1.1 & 1.4 & 2.1 & 2.6 \\
         & ACRec & {\bf 5.5} & {\bf 12.3} & {\bf 15.6} & {\bf 25.6} & {\bf 29.4} & {\bf 9.1} & {\bf 10.1} & {\bf 12.4} & {\bf 13.0} \\
         \hline
    A + AC & ACRec & 1.1 & 2.5 & 2.9 & 4.9 & 6.7 & 1.8 & 1.9 & 2.3 & 2.6 \\
         \hline
    \end{tabular}
    \caption{Hit Ratio (HR in \%) and Normalized Discounted Cumulative Gain (NDCG in \%) metrics when ranking all 143K items in the dataset.  Results shown for A + AC + C (1,000 users) and A + AC (554 users)}
    \label{tab:results-all-items}
\end{table*}

\begin{table}[t]
    \centering
    \small
    \begin{tabular}{l|cc}
    \hline
    System & A + AC + C & A + AC \\
    \hline
    SASRec & 18.6 & 20.0 \\
    BERT4Rec & 15.1 & 16.2 \\
    UniSRec & 11.9 & 11.7 \\
    RecFormer & 44.2 & 40.0 \\
    LLaRA & 20.3 & 19.4 \\
    LLaRA-AC & 40.9 & 37.6 \\
    iLoRA & 20.9 & 21.2 \\
    iLoRA-AC & 40.3 & 42.7 \\
    ACRec & {\bf 67.6} & {\bf 43.4} \\
    \hline
    \end{tabular}
    \caption{Top-1 accuracy (\%) when ranking 19 randomly sampled candidate items + the 1 ground truth item.}
    \label{tab:results-20-items}
\end{table}

According to validation experiments reported in Appendix~\ref{sec:validation-appendix}, the best ACRec architecture has 2 hidden layers in the fully connected neural network (FCN) module and benefits from the incorporation of Cosine as a feature. We compare ACRec against three groups of sequential baselines: (1) ID-based models SASRec \cite{sasrec} and BERT4Rec \cite{bert4rec}, which use only item IDs; (2) Language Model-based methods UniSRec \cite{unisrec} and RecFormer \cite{recformer}, which incorporate item descriptions; and (3) LLM-based hybrids LLaRA \cite{llara} and iLoRA \cite{ilora}, which combine item titles with ID-based architectures. Groups 1 and 2 systems can be used at test time to rank the full item set, while Group 3 systems use item titles in prompts and thus can rank only a limited number of items (20) due to prompt size constraints. Thus, we evaluate under two settings: (i) all-item ranking (excluding previously consumed items) for Groups 1 and 2, and (ii) 20-item ranking (1 ground-truth + 19 random negatives) for all models. 


To the best of our knowledge, no existing recommender system directly incorporates AC statements as input. Nevertheless, we modified the prompts of Group 3 models to enable specification of AC statements; we refer to these versions as LLaRA-AC and iLoRA-AC. Attempts to incorporate AC descriptions as a special book description in the inputs of UniSRec and RecFormer led to a worsening of their performance. More details about the baselines and their modified versions are provided in Appendix~\ref{sec:baselines}. 

Table~\ref{tab:results-all-items} presents the performance of ACRec and baseline models in the all-item ranking setting. ACRec consistently outperforms all baselines by a large margin across all ranks and metrics. When provided with the full AC description (A + AC + C), ACRec achieves a HR@10 of 15.6\% and NDCG@10 of 10.1\%, whereas the strongest competing baseline, RecFormer, achieves only 2.8\% and 1.4\%, respectively. These results highlight the importance of explicitly modeling user-specified affective-cognitive preferences in the recommendation process. Among the baselines, ID-based models such as SASRec and BERT4Rec perform the worst, indicating that ID-only representations are insufficient for capturing nuanced affective needs. 

We also evaluate ACRec using only the AC statements that have affective content (A + AC). The resulting drop in performance, e.g., from 15.6\% to 2.9\% in HR@10, is expected, as the objective content in C statements adds substantial distinguishing information. Note that, overall, all results are conservative in that it is likely that in reality the ground truth book (that the user actually read) is not the only book satisfying the AC description.

Table~\ref{tab:results-20-items} reports Top-1 accuracy when ranking the ground-truth item among 20 candidates, a setting that enables comparison with LLM-based hybrid models such as LLaRA and iLoRA. Once again, ACRec outperforms all other models by a large margin, achieving an accuracy of 67.6\% when using the full AC description (A + AC + C),  RecFormer, the strongest among non-LLM baselines, reaches only 44.2\%, while the enhanced LLM variants LLaRA-AC and iLoRA-AC reach 40.9\% and 42.7\%, respectively. Notably, incorporating AC statements into these LLM-based models (LLaRA-AC and iLoRA-AC) leads to substantial improvements, doubling their Top-1 accuracy to 40.9\% and 42.7\%, respectively. This confirms that affective-cognitive information is beneficial not just for ACRec, but also for other recommender models.

When using only the A + AC descriptions, ACRec’s performance drops to 43.4\%, still outperforming RecFormer (40.0\%). The drop from 67.6\% to 43.4\% highlights that when users provide richer, more informative AC statements, including both how they want to feel and what kinds of content or structure they seek, ACRec is better able to align those preferences with appropriate recommendations.

Error analysis of the ACRec model revealed several reasons for the positive sample being ranked out the top 10, such as when the AC descriptions are overly generic, e.g.,  \textit{“The book will be so much fun to read”}. Another issue arises when AC descriptions contain proper nouns. For instance, one AC description includes the word \textit{Australia}. Although the description of the ground-truth book also mentions {\it Australia}, the model tends to recommend books where {\it Australia} appears more frequently or prominently in their original or extended descriptions. Finally, it is important to note that the experimental results are conservative: the fact that the ground truth positive book is not ranked at the top does not necessarily mean that books ranked above it would not induce the affective-cognitive states mentioned in the AC description.

\section{Related Work}
\label{sec:related}

In affective recommender systems, \citet{8675268} use a neural attention model to incorporate sentiment from user reviews. \citet{mizgajski2019affective} address challenges in emotion-based news recommendation by introducing emotion collection methods. Affective signals have also been used in short film and movie recommendations via collaborative filtering and context mining \cite{10.1145/2700171.2791042,10.1145/3106426.3106535}. EmoWare \cite{8691425} leverages users' non-verbal emotional cues for context-aware video recommendations. \citet{10.3389/fnins.2022.984404} survey affective video recommenders, while \citet{POLIGNANO2021114382} propose an affective coherence model to capture emotional transitions between items. Literature reviews highlight growing interest in this area across domains, including education \cite{10.1016/j.cosrev.2021.100377, KATARYA2016182}. However, these systems primarily rely on users’ {\it past} affective responses to recommend items they may like. In contrast, our proposed ACRec framework enables users to specify a broad range of desired affective states and recommends items likely to induce those states in the {\it future}.

Beyond-accuracy objectives, such as diversity, novelty, unexpectedness, and serendipity, have gained significant attention in recommender systems. \citet{ijcai2019p380} enhanced diversity in sequential recommendations, while the system of \citet{10.1145/3404855} identified surprising or unexpected items. How diversity, serendipity, novelty, and coverage can enhance user satisfaction and engagement in recommender systems has been discussed along accuracy in \cite{10.1145/2926720}. 
Serendipity is explored extensively in other works, such as \cite{10.1145/2124295.2124300,4626624,10.1145/3604915.3608851,10.1145/3605145,KOTKOV2016180}. Overall,  the novelty, unexpectedness, and serendipity criteria emphasized in the systems cited above can be seen as special cases in the wide range of affective-cognitive states addressed by our system, e.g. the user's AC description can stipulate {\it "I would like a book that surprises me"} or {\it "a book that is unlike any I have read so far"}.

In conversational recommender systems (CRSs), \citet{10.1145/3340531.3412098} infer item-level preferences from historical interactions and attribute-level preferences from dialogue. \citet{10.1145/3336191.3371769} propose an EAR (estimation-action-reflection) framework that leverages item attributes without incorporating long-term interaction history. Knowledge graph-based methods improve recommendation accuracy by modeling users' temporal conversations \cite{10.1145/3394486.3403143,10.1145/3240323.3240338,10.1145/3534678.3539382,10.1145/3677376,10.1145/3627043.3659565}. The affective-cognitive dimension remains underexplored; \citet{zhang2024towards} introduce empathy by detecting emotions in conversations to align recommendations with users’ affective states, identifying nine primary emotions to extract local emotion-aware entities. While a promising step, such approaches are limited in capturing the broader range of user experiences and emotions. In contrast, our work enables users to express and select arbitrarily fine-grained affective states beyond attitudes toward objective features.

\section{Conclusion and Future Work}
\label{sec:conclusion}

We introduced a novel text-based recommendation setting where users express affective-cognitive states they would like to experience upon reading a book. Expressions of AC states were mined from real book reviews, and then used for the training and evaluation of recommendation architecture exploiting short-term and long-term user preferences implicit in their readings, reviews, and ratings histories. Experimental evaluations show that the proposed approach is effective at identifying books that match user's expressed affective preferences and compares favorably with state-of-the-art recommender systems. Future work includes combining ID-based with text-based information in the sequential modeling of reading histories, and building an easy to navigate ontology of AC statements complete with a graphical user interface.


\section*{Acknowledgments}

We thank Khushi Patel for her help with crafting part of the prompt for extracting AC statements from reviews. The Microsoft AFMR program provided generous Azure credits for the LLM experiments.

\section*{Limitations}
\label{sec:limitations}

Although expressions of affective-cognitive preferences can be utilized in other domains such as movies, in this paper we only investigated their utility through the lens of book recommendations. Furthermore, the experimental results were obtained on users having a sufficiently large reading history to enable learning of short-term and long-term user preferences. As such, performance for all tested systems is likely to be lower when text data is limited or unavailable, such as in cold-start scenarios. As such, in future work we plan to investigate alternative ways of modeling users' affective states when the user history is limited. We also plan to explore the applicability of affective-cognitive descriptions to the movie recommendation domain.

\section*{Ethical Considerations} 

The AC dataset presented in this paper has been derived from a publicly available dataset that has been utilized in previous studies. This dataset does not contain any sensitive user information, ensuring compliance with ethical standards. Nevertheless, during the generation of affective-cognitive descriptions from original reviews, we encountered instances where the LLM did not allow processing some reviews, displaying the error: "violated content policies due to the presence of adult or hate content." Such reviews were omitted from the dataset.

\bibliography{arxiv25}

\begin{thebibliography}{49}
\providecommand{\natexlab}[1]{#1}

\bibitem[{Anelli et~al.(2018)Anelli, Basile, Bridge, Di~Noia, Lops, Musto, Narducci, and Zanker}]{10.1145/3240323.3240338}
Vito~Walter Anelli, Pierpaolo Basile, Derek Bridge, Tommaso Di~Noia, Pasquale Lops, Cataldo Musto, Fedelucio Narducci, and Markus Zanker. 2018.
\newblock \href {https://doi.org/10.1145/3240323.3240338} {Knowledge-aware and conversational recommender systems}.
\newblock In \emph{Proceedings of the 12th ACM Conference on Recommender Systems}, RecSys '18, page 521–522, New York, NY, USA. Association for Computing Machinery.

\bibitem[{Da’u and Salim(2019)}]{8675268}
Aminu Da’u and Naomie Salim. 2019.
\newblock \href {https://doi.org/10.1109/ACCESS.2019.2907729} {Sentiment-aware deep recommender system with neural attention networks}.
\newblock \emph{IEEE Access}, 7:45472--45484.

\bibitem[{Fu et~al.(2023)Fu, Niu, and Maher}]{10.1145/3605145}
Zhe Fu, Xi~Niu, and Mary~Lou Maher. 2023.
\newblock \href {https://doi.org/10.1145/3605145} {Deep learning models for serendipity recommendations: A survey and new perspectives}.
\newblock \emph{ACM Comput. Surv.}, 56(1).

\bibitem[{Goldie(2004)}]{goldie_life_2004}
Peter Goldie. 2004.
\newblock \href {https://doi.org/10.1177/0539018404047705} {The life of the mind: commentary on “{Emotions} in everyday life”}.
\newblock \emph{Social Science Information}, 43(4):591--598.
\newblock Publisher: SAGE Publications Ltd.

\bibitem[{Hasan and Bunescu(2023)}]{10.1145/3604915.3608851}
Tonmoy Hasan and Razvan Bunescu. 2023.
\newblock \href {https://doi.org/10.1145/3604915.3608851} {Topic-level bayesian surprise and serendipity for recommender systems}.
\newblock In \emph{Proceedings of the 17th ACM Conference on Recommender Systems}, RecSys '23, page 933–939, New York, NY, USA. Association for Computing Machinery.

\bibitem[{Hayati et~al.(2020)Hayati, Kang, Zhu, Shi, and Yu}]{hayati_inspired_2020}
Shirley~Anugrah Hayati, Dongyeop Kang, Qingxiaoyang Zhu, Weiyan Shi, and Zhou Yu. 2020.
\newblock \href {https://doi.org/10.18653/v1/2020.emnlp-main.654} {{INSPIRED}: {Toward} {Sociable} {Recommendation} {Dialog} {Systems}}.
\newblock In \emph{Proceedings of the 2020 {Conference} on {Empirical} {Methods} in {Natural} {Language} {Processing} ({EMNLP})}, pages 8142--8152, Online. Association for Computational Linguistics.

\bibitem[{He et~al.(2023)He, Xie, Jha, Steck, Liang, Feng, Majumder, Kallus, and Mcauley}]{he_large_2023}
Zhankui He, Zhouhang Xie, Rahul Jha, Harald Steck, Dawen Liang, Yesu Feng, Bodhisattwa~Prasad Majumder, Nathan Kallus, and Julian Mcauley. 2023.
\newblock \href {https://doi.org/10.1145/3583780.3614949} {Large {Language} {Models} as {Zero}-{Shot} {Conversational} {Recommenders}}.
\newblock In \emph{Proceedings of the 32nd {ACM} {International} {Conference} on {Information} and {Knowledge} {Management}}, {CIKM} '23, pages 720--730, New York, NY, USA. Association for Computing Machinery.

\bibitem[{Hou et~al.(2022)Hou, Mu, Zhao, Li, Ding, and Wen}]{unisrec}
Yupeng Hou, Shanlei Mu, Wayne~Xin Zhao, Yaliang Li, Bolin Ding, and Ji-Rong Wen. 2022.
\newblock \href {https://doi.org/10.1145/3534678.3539381} {Towards universal sequence representation learning for recommender systems}.
\newblock In \emph{Proceedings of the 28th ACM SIGKDD Conference on Knowledge Discovery and Data Mining}, KDD '22, page 585–593, New York, NY, USA. Association for Computing Machinery.

\bibitem[{Iaquinta et~al.(2008)Iaquinta, de~Gemmis, Lops, Semeraro, Filannino, and Molino}]{4626624}
Leo Iaquinta, Marco de~Gemmis, Pasquale Lops, Giovanni Semeraro, Michele Filannino, and Piero Molino. 2008.
\newblock \href {https://doi.org/10.1109/HIS.2008.25} {Introducing serendipity in a content-based recommender system}.
\newblock In \emph{2008 Eighth International Conference on Hybrid Intelligent Systems}, pages 168--173.

\bibitem[{Kaminskas and Bridge(2016)}]{10.1145/2926720}
Marius Kaminskas and Derek Bridge. 2016.
\newblock \href {https://doi.org/10.1145/2926720} {Diversity, serendipity, novelty, and coverage: A survey and empirical analysis of beyond-accuracy objectives in recommender systems}.
\newblock \emph{ACM Trans. Interact. Intell. Syst.}, 7(1).

\bibitem[{Kang and McAuley(2018)}]{sasrec}
Wang-Cheng Kang and Julian McAuley. 2018.
\newblock \href {https://doi.org/10.1109/ICDM.2018.00035} {Self-attentive sequential recommendation}.
\newblock In \emph{2018 IEEE International Conference on Data Mining (ICDM)}, pages 197--206.

\bibitem[{Katarya and Verma(2016)}]{KATARYA2016182}
Rahul Katarya and Om~Prakash Verma. 2016.
\newblock \href {https://doi.org/10.1016/j.physa.2016.05.046} {Recent developments in affective recommender systems}.
\newblock \emph{Physica A: Statistical Mechanics and its Applications}, 461:182--190.

\bibitem[{Kim et~al.(2019)Kim, Kim, Park, and Yu}]{ijcai2019p380}
Yejin Kim, Kwangseob Kim, Chanyoung Park, and Hwanjo Yu. 2019.
\newblock \href {https://doi.org/10.24963/ijcai.2019/380} {Sequential and diverse recommendation with long tail}.
\newblock In \emph{Proceedings of the Twenty-Eighth International Joint Conference on Artificial Intelligence, {IJCAI-19}}, pages 2740--2746. International Joint Conferences on Artificial Intelligence Organization.

\bibitem[{Kingma and Ba(2014)}]{kingma2014adam}
Diederik~P Kingma and Jimmy Ba. 2014.
\newblock Adam: A method for stochastic optimization.
\newblock \emph{arXiv preprint arXiv:1412.6980}.

\bibitem[{Kong et~al.(2024)Kong, Wu, Zhang, Sheng, Lin, Wang, and He}]{ilora}
Xiaoyu Kong, Jiancan Wu, An~Zhang, Leheng Sheng, Hui Lin, Xiang Wang, and Xiangnan He. 2024.
\newblock \href {https://proceedings.neurips.cc/paper_files/paper/2024/file/cd476d01692c508ddf1cb43c6279a704-Paper-Conference.pdf} {Customizing language models with instance-wise lora for sequential recommendation}.
\newblock In \emph{Advances in Neural Information Processing Systems}, volume~37, pages 113072--113095. Curran Associates, Inc.

\bibitem[{Kotkov et~al.(2016)Kotkov, Wang, and Veijalainen}]{KOTKOV2016180}
Denis Kotkov, Shuaiqiang Wang, and Jari Veijalainen. 2016.
\newblock \href {https://doi.org/10.1016/j.knosys.2016.08.014} {A survey of serendipity in recommender systems}.
\newblock \emph{Knowledge-Based Systems}, 111:180--192.

\bibitem[{Lei et~al.(2020)Lei, He, Miao, Wu, Hong, Kan, and Chua}]{10.1145/3336191.3371769}
Wenqiang Lei, Xiangnan He, Yisong Miao, Qingyun Wu, Richang Hong, Min-Yen Kan, and Tat-Seng Chua. 2020.
\newblock \href {https://doi.org/10.1145/3336191.3371769} {Estimation-action-reflection: Towards deep interaction between conversational and recommender systems}.
\newblock In \emph{Proceedings of the 13th International Conference on Web Search and Data Mining}, WSDM '20, page 304–312, New York, NY, USA. Association for Computing Machinery.

\bibitem[{Li et~al.(2023)Li, Wang, Li, Fu, Shen, Shang, and McAuley}]{recformer}
Jiacheng Li, Ming Wang, Jin Li, Jinmiao Fu, Xin Shen, Jingbo Shang, and Julian McAuley. 2023.
\newblock \href {https://doi.org/10.1145/3580305.3599519} {Text is all you need: Learning language representations for sequential recommendation}.
\newblock In \emph{Proceedings of the 29th ACM SIGKDD Conference on Knowledge Discovery and Data Mining}, KDD '23, page 1258–1267, New York, NY, USA. Association for Computing Machinery.

\bibitem[{Li and Tuzhilin(2020)}]{10.1145/3404855}
Pan Li and Alexander Tuzhilin. 2020.
\newblock \href {https://doi.org/10.1145/3404855} {Latent unexpected recommendations}.
\newblock \emph{ACM Trans. Intell. Syst. Technol.}, 11(6).

\bibitem[{Li et~al.(2018)Li, Ebrahimi~Kahou, Schulz, Michalski, Charlin, and Pal}]{li_towards_2018}
Raymond Li, Samira Ebrahimi~Kahou, Hannes Schulz, Vincent Michalski, Laurent Charlin, and Chris Pal. 2018.
\newblock \href {https://papers.nips.cc/paper_files/paper/2018/hash/800de15c79c8d840f4e78d3af937d4d4-Abstract.html} {Towards {Deep} {Conversational} {Recommendations}}.
\newblock In \emph{Advances in {Neural} {Information} {Processing} {Systems}}, volume~31. Curran Associates, Inc.

\bibitem[{Lian et~al.(2024)Lian, Lei, Huang, Yao, Xu, and Xie}]{lian_recai_2024}
Jianxun Lian, Yuxuan Lei, Xu~Huang, Jing Yao, Wei Xu, and Xing Xie. 2024.
\newblock \href {https://doi.org/10.1145/3589335.3651242} {{RecAI}: {Leveraging} {Large} {Language} {Models} for {Next}-{Generation} {Recommender} {Systems}}.
\newblock In \emph{Companion {Proceedings} of the {ACM} {Web} {Conference} 2024}, {WWW} '24, pages 1031--1034, New York, NY, USA. Association for Computing Machinery.

\bibitem[{Liao et~al.(2024)Liao, Li, Yang, Wu, Yuan, Wang, and He}]{llara}
Jiayi Liao, Sihang Li, Zhengyi Yang, Jiancan Wu, Yancheng Yuan, Xiang Wang, and Xiangnan He. 2024.
\newblock \href {https://doi.org/10.1145/3626772.3657690} {Llara: Large language-recommendation assistant}.
\newblock In \emph{Proceedings of the 47th International ACM SIGIR Conference on Research and Development in Information Retrieval}, SIGIR '24, page 1785–1795, New York, NY, USA. Association for Computing Machinery.

\bibitem[{Mizgajski and Morzy(2019)}]{mizgajski2019affective}
Jan Mizgajski and Miko{\l}aj Morzy. 2019.
\newblock Affective recommender systems in online news industry: how emotions influence reading choices.
\newblock \emph{User Modeling and User-Adapted Interaction}, 29(2):345--379.

\bibitem[{Orellana-Rodriguez et~al.(2015)Orellana-Rodriguez, Diaz-Aviles, and Nejdl}]{10.1145/2700171.2791042}
Claudia Orellana-Rodriguez, Ernesto Diaz-Aviles, and Wolfgang Nejdl. 2015.
\newblock \href {https://doi.org/10.1145/2700171.2791042} {Mining affective context in short films for emotion-aware recommendation}.
\newblock In \emph{Proceedings of the 26th ACM Conference on Hypertext \& Social Media}, HT '15, page 185–194, New York, NY, USA. Association for Computing Machinery.

\bibitem[{Park et~al.(2012)Park, Kim, Choi, and Kim}]{park:esa12}
Deuk~Hee Park, Hyea~Kyeong Kim, Il~Young Choi, and Jae~Kyeong Kim. 2012.
\newblock \href {https://doi.org/10.1016/j.eswa.2012.02.038} {A literature review and classification of recommender systems research}.
\newblock \emph{Expert Systems with Applications}, 39(11):10059--10072.

\bibitem[{Penha and Hauff(2020)}]{penha_what_2020}
Gustavo Penha and Claudia Hauff. 2020.
\newblock \href {https://doi.org/10.1145/3383313.3412249} {What does {BERT} know about books, movies and music? {Probing} {BERT} for {Conversational} {Recommendation}}.
\newblock In \emph{Proceedings of the 14th {ACM} {Conference} on {Recommender} {Systems}}, {RecSys} '20, pages 388--397, New York, NY, USA. Association for Computing Machinery.

\bibitem[{Petruzzelli et~al.(2024)Petruzzelli, Martina, Spillo, Musto, De~Gemmis, Lops, and Semeraro}]{10.1145/3627043.3659565}
Alessandro Petruzzelli, Alessandro Francesco~Maria Martina, Giuseppe Spillo, Cataldo Musto, Marco De~Gemmis, Pasquale Lops, and Giovanni Semeraro. 2024.
\newblock \href {https://doi.org/10.1145/3627043.3659565} {Improving transformer-based sequential conversational recommendations through knowledge graph embeddings}.
\newblock In \emph{Proceedings of the 32nd ACM Conference on User Modeling, Adaptation and Personalization}, UMAP '24, page 172–182, New York, NY, USA. Association for Computing Machinery.

\bibitem[{Polignano et~al.(2021)Polignano, Narducci, {de Gemmis}, and Semeraro}]{POLIGNANO2021114382}
Marco Polignano, Fedelucio Narducci, Marco {de Gemmis}, and Giovanni Semeraro. 2021.
\newblock \href {https://doi.org/10.1016/j.eswa.2020.114382} {Towards emotion-aware recommender systems: an affective coherence model based on emotion-driven behaviors}.
\newblock \emph{Expert Systems with Applications}, 170:114382.

\bibitem[{Rendle et~al.(2009)Rendle, Freudenthaler, Gantner, and Schmidt-Thieme}]{10.5555/1795114.1795167}
Steffen Rendle, Christoph Freudenthaler, Zeno Gantner, and Lars Schmidt-Thieme. 2009.
\newblock Bpr: Bayesian personalized ranking from implicit feedback.
\newblock In \emph{Proceedings of the Twenty-Fifth Conference on Uncertainty in Artificial Intelligence}, UAI '09, page 452–461, Arlington, Virginia, USA. AUAI Press.

\bibitem[{Resnick and Varian(1997)}]{resnick:acm97}
Paul Resnick and Hal~R. Varian. 1997.
\newblock \href {https://doi.org/10.1145/245108.245121} {Recommender systems}.
\newblock \emph{Communications of the ACM}, 40(3):56–58.

\bibitem[{Sacharin et~al.(2012)Sacharin, K., and R.}]{gew:12}
V.~Sacharin, Schlegel K., and Scherer~K. R. 2012.
\newblock \href {https://www.unige.ch/cisa/files/4514/6720/4016/Geneva_Emotion_Wheel_Rating_Study_Report_2012_08_11_2.0.pdf} {{Geneva Emotion Wheel} rating study}.
\newblock Unpublished report.
\newblock University of Geneva: Swiss Center for Affective Studies.

\bibitem[{Salazar et~al.(2021)Salazar, Aguilar, Monsalve-Pulido, and Montoya}]{10.1016/j.cosrev.2021.100377}
Camilo Salazar, Jose Aguilar, Juli\'{a}n Monsalve-Pulido, and Edwin Montoya. 2021.
\newblock \href {https://doi.org/10.1016/j.cosrev.2021.100377} {Affective recommender systems in the educational field. a systematic literature review}.
\newblock \emph{Comput. Sci. Rev.}, 40(C).

\bibitem[{Scherer(2005)}]{scherer_what_2005}
Klaus~R. Scherer. 2005.
\newblock \href {https://doi.org/10.1177/0539018405058216} {What are emotions? {And} how can they be measured?}
\newblock \emph{Social Science Information}, 44(4):695--729.
\newblock Publisher: SAGE Publications Ltd.

\bibitem[{Sturua et~al.(2024)Sturua, Mohr, Akram, G{\"u}nther, Wang, Krimmel, Wang, Mastrapas, Koukounas, Wang et~al.}]{sturua2024jina}
Saba Sturua, Isabelle Mohr, Mohammad~Kalim Akram, Michael G{\"u}nther, Bo~Wang, Markus Krimmel, Feng Wang, Georgios Mastrapas, Andreas Koukounas, Nan Wang, et~al. 2024.
\newblock jina-embeddings-v3: Multilingual embeddings with task lora.
\newblock \emph{arXiv preprint arXiv:2409.10173}.

\bibitem[{Sun et~al.(2019)Sun, Liu, Wu, Pei, Lin, Ou, and Jiang}]{bert4rec}
Fei Sun, Jun Liu, Jian Wu, Changhua Pei, Xiao Lin, Wenwu Ou, and Peng Jiang. 2019.
\newblock \href {https://doi.org/10.1145/3357384.3357895} {Bert4rec: Sequential recommendation with bidirectional encoder representations from transformer}.
\newblock In \emph{Proceedings of the 28th ACM International Conference on Information and Knowledge Management}, CIKM '19, page 1441–1450, New York, NY, USA. Association for Computing Machinery.

\bibitem[{Thompson et~al.(2004)Thompson, Göker, and Langley}]{thompson_personalized_2004}
Cynthia~A. Thompson, Mehmet~H. Göker, and Pat Langley. 2004.
\newblock A personalized system for conversational recommendations.
\newblock \emph{J. Artif. Int. Res.}, 21(1):393--428.

\bibitem[{Tripathi et~al.(2019)Tripathi, Ashwin, and Guddeti}]{8691425}
Abhishek Tripathi, T.~S. Ashwin, and Ram Mohana~Reddy Guddeti. 2019.
\newblock \href {https://doi.org/10.1109/ACCESS.2019.2911235} {Emoware: A context-aware framework for personalized video recommendation using affective video sequences}.
\newblock \emph{IEEE Access}, 7:51185--51200.

\bibitem[{Vaswani et~al.(2017)Vaswani, Shazeer, Parmar, Uszkoreit, Jones, Gomez, Kaiser, and Polosukhin}]{NIPS2017_3f5ee243}
Ashish Vaswani, Noam Shazeer, Niki Parmar, Jakob Uszkoreit, Llion Jones, Aidan~N Gomez, \L~ukasz Kaiser, and Illia Polosukhin. 2017.
\newblock \href {https://proceedings.neurips.cc/paper_files/paper/2017/file/3f5ee243547dee91fbd053c1c4a845aa-Paper.pdf} {Attention is all you need}.
\newblock In \emph{Advances in Neural Information Processing Systems}, volume~30. Curran Associates, Inc.

\bibitem[{Wan and McAuley(2018{\natexlab{a}})}]{wan2018goodreads}
Mengting Wan and Julian McAuley. 2018{\natexlab{a}}.
\newblock \href {https://doi.org/10.1145/3240323.3240369} {Item recommendation on monotonic behavior chains}.
\newblock In \emph{Proceedings of the 12th ACM Conference on Recommender Systems}, RecSys '18, page 86–94, New York, NY, USA. Association for Computing Machinery.

\bibitem[{Wan and McAuley(2018{\natexlab{b}})}]{10.1145/3240323.3240369}
Mengting Wan and Julian McAuley. 2018{\natexlab{b}}.
\newblock \href {https://doi.org/10.1145/3240323.3240369} {Item recommendation on monotonic behavior chains}.
\newblock In \emph{Proceedings of the 12th ACM Conference on Recommender Systems}, RecSys '18, page 86–94, New York, NY, USA. Association for Computing Machinery.

\bibitem[{Wang and Zhao(2022)}]{10.3389/fnins.2022.984404}
Dandan Wang and Xiaoming Zhao. 2022.
\newblock \href {https://doi.org/10.3389/fnins.2022.984404} {Affective video recommender systems: A survey}.
\newblock \emph{Frontiers in Neuroscience}, 16.

\bibitem[{Wang et~al.(2022)Wang, Zhou, Wen, and Zhao}]{10.1145/3534678.3539382}
Xiaolei Wang, Kun Zhou, Ji-Rong Wen, and Wayne~Xin Zhao. 2022.
\newblock \href {https://doi.org/10.1145/3534678.3539382} {Towards unified conversational recommender systems via knowledge-enhanced prompt learning}.
\newblock In \emph{Proceedings of the 28th ACM SIGKDD Conference on Knowledge Discovery and Data Mining}, KDD '22, page 1929–1937, New York, NY, USA. Association for Computing Machinery.

\bibitem[{Zhang et~al.(2024)Zhang, Xie, Lyu, Xin, Ren, Liang, Zhang, Kang, de~Rijke, and Ren}]{zhang2024towards}
Xiaoyu Zhang, Ruobing Xie, Yougang Lyu, Xin Xin, Pengjie Ren, Mingfei Liang, Bo~Zhang, Zhanhui Kang, Maarten de~Rijke, and Zhaochun Ren. 2024.
\newblock Towards empathetic conversational recommender systems.
\newblock \emph{arXiv preprint arXiv:2409.10527}.

\bibitem[{Zhang et~al.(2012)Zhang, S\'{e}aghdha, Quercia, and Jambor}]{10.1145/2124295.2124300}
Yuan~Cao Zhang, Diarmuid~\'{O} S\'{e}aghdha, Daniele Quercia, and Tamas Jambor. 2012.
\newblock \href {https://doi.org/10.1145/2124295.2124300} {Auralist: introducing serendipity into music recommendation}.
\newblock In \emph{Proceedings of the Fifth ACM International Conference on Web Search and Data Mining}, WSDM '12, page 13–22, New York, NY, USA. Association for Computing Machinery.

\bibitem[{Zhao et~al.(2024)Zhao, Fan, Li, Liu, Mei, Wang, Wen, Wang, Zhao, Tang, and Li}]{zhao_recommender_2024}
Zihuai Zhao, Wenqi Fan, Jiatong Li, Yunqing Liu, Xiaowei Mei, Yiqi Wang, Zhen Wen, Fei Wang, Xiangyu Zhao, Jiliang Tang, and Qing Li. 2024.
\newblock \href {https://doi.org/10.1109/TKDE.2024.3392335} {Recommender {Systems} in the {Era} of {Large} {Language} {Models} ({LLMs})}.
\newblock \emph{IEEE Transactions on Knowledge and Data Engineering}, (01):1--20.
\newblock Publisher: IEEE Computer Society.

\bibitem[{Zheng(2017)}]{10.1145/3106426.3106535}
Yong Zheng. 2017.
\newblock \href {https://doi.org/10.1145/3106426.3106535} {Affective prediction by collaborative chains in movie recommendation}.
\newblock In \emph{Proceedings of the International Conference on Web Intelligence}, WI '17, page 815–822, New York, NY, USA. Association for Computing Machinery.

\bibitem[{Zhou et~al.(2020{\natexlab{a}})Zhou, Zhao, Bian, Zhou, Wen, and Yu}]{10.1145/3394486.3403143}
Kun Zhou, Wayne~Xin Zhao, Shuqing Bian, Yuanhang Zhou, Ji-Rong Wen, and Jingsong Yu. 2020{\natexlab{a}}.
\newblock \href {https://doi.org/10.1145/3394486.3403143} {Improving conversational recommender systems via knowledge graph based semantic fusion}.
\newblock In \emph{Proceedings of the 26th ACM SIGKDD International Conference on Knowledge Discovery \& Data Mining}, KDD '20, page 1006–1014, New York, NY, USA. Association for Computing Machinery.

\bibitem[{Zhou et~al.(2020{\natexlab{b}})Zhou, Zhao, Wang, Wang, Zhang, Wang, and Wen}]{10.1145/3340531.3412098}
Kun Zhou, Wayne~Xin Zhao, Hui Wang, Sirui Wang, Fuzheng Zhang, Zhongyuan Wang, and Ji-Rong Wen. 2020{\natexlab{b}}.
\newblock \href {https://doi.org/10.1145/3340531.3412098} {Leveraging historical interaction data for improving conversational recommender system}.
\newblock In \emph{Proceedings of the 29th ACM International Conference on Information \& Knowledge Management}, CIKM '20, page 2349–2352, New York, NY, USA. Association for Computing Machinery.

\bibitem[{Zou et~al.(2024)Zou, Sun, Long, and Kanoulas}]{10.1145/3677376}
Jie Zou, Aixin Sun, Cheng Long, and Evangelos Kanoulas. 2024.
\newblock \href {https://doi.org/10.1145/3677376} {Knowledge-enhanced conversational recommendation via transformer-based sequential modelling}.
\newblock \emph{ACM Trans. Inf. Syst.}
\newblock Just Accepted.

\end{thebibliography}

\appendix

\section{Generation of AC statements}
\label{sec:dataset_appendix}

Due to size constraints, the prompts and the few-shot examples used with the GPT-4o  model in order to generate AC statements are made available at the following link\footnote{\url{https://github.com/Ton-moy/ACRec_Taxonomy}}. The following criteria were applied to manually evaluate the quality of the GPT-4o-generated AC statements:
\begin{itemize}
    \item Does the generated AC text exclude identifiable book information, e.g. the title?
    \item Does the generated AC text exclude identifiable author information, e.g. the author's name?
    \item Does the generated AC text exclude identifiable character information, e.g. the character's name?
    \item Does the generated AC text omit sentences directly taken from the book?
    \item Does the generated AC text exclude sentences with negative or partially negative connotation towards the book?
    \item Does the generated AC text exclude chapter numbers, chapter names, and page number?
\end{itemize}

\section{Validation Experiments}
\label{sec:validation-appendix}

We used the validation time steps to evaluate 4 models: {\bf Cosine}, {\bf FCN}, {\bf ACRec}, and {\bf ACRecCos}. To evaluate each model's performance, we use HitRate@10 and NDCG@10: HR@10 is the fraction of times the ground-truth positive book appears among top-10 recommended books, whereas NDCG@10 is a position aware metric calculated as the inverse of $\log_2(r_b+1)$, where $r_b$ is the rank of the ground-truth positive book $b$ if among the top-10 items, otherwise 0.

\begin{table}[h]
    \centering
    \small
    \begin{tabular}{lcc}
        \hline
        Model       & HR@10 & NDCG@10 \\
        \hline
        {\bf Cosine}         & {\bf 81.6}  & {\bf 66.2}   \\
        \hline
        FCN-0       & 39.0  & 25.1    \\
        FCN-1       & 67.6  & 42.7    \\
        FCN-2       & 75.8  & 49.1    \\
        {\bf FCN-3}       & {\bf 77.2}  & {\bf 52.1}    \\
        FCN-4       & 75.6  & 51.1    \\
        \hline
        ACRec-1     & 81.7  & 54.3    \\
        {\bf ACRec-2}     & {\bf 86.5}  & {\bf 60.8}    \\
        ACRec-3     & 86.4  & 60.3    \\
        \hline
        {\bf ACRec-2+Cos} & {\bf 88.3}  & {\bf 67.1}    \\
        \hline
    \end{tabular}
    \caption{Validation performance over 1,000 users.}
    \label{tab:ae-eval}
\end{table}

In the Cosine model, we create a combined book description by concatenating the books' original and extended descriptions. For each test sample, we calculate the cosine similarity between the Jina embedding of the AC description $\mathbf{ac}$ and the Jina embeddings of the combined book description $\mathbf{d}_b$. The 101 test samples are then ranked based on their cosine similarity score.

In the FCN model, the Jina embeddings of the original book description, extended book description, and AC description are ran through three separate linear projection layers, followed by a ramp activation function, reducing their dimensionality to 42, 43 and 43, respectively. These projected embeddings are then concatenated to form a 128-dinmensional vector, which is then passed through $n$ hidden layers before a final linear layer produces the recommendation score. During validation we experimented with $n \in \{0, 1, 2, 3, 4\}$, with the respective models being called FCN-$n$. Based on the validation performance showed in Table~\ref{tab:ae-eval}, we selected FCN-3 as the final FCN model for testing.

The base ACRec model is described in Section~\ref{sec:acrec-framework}. During validation we considered three variants ACRec-1, ACRec-2, and ACRec-3, that were using 1, 2, or 3 hidden layers, respectively, in the fully connected network computing the recommendation score. Based on the validation performance showed in Table~\ref{tab:ae-eval}, we selected ACRec-2 as the base ACRec model for testing. The short-term history was set to the last 30 items, based on the validation results shown in Table~\ref{tab:ae-m}. The higher performance obtained by the ACRec-2 model in Table~\ref{tab:ae-eval} vs. Table~\ref{tab:ae-m} shows the impact of utilizing long-term preferences in the model. Additional hyperparameters and their tuning procedures are described in Appendix~\ref{sec:tuning}.

According to the validation results Table~\ref{tab:ae-eval}, Cosine achieves better NDCG than the best ACRec model. Therefore, we created a combined model ACRecCos that uses the cosine similarity as additional input to the fully connected network computing the recommendation score, as shown through the dashed line in Figure~\ref{fig:framework}.
Correspondingly, the recommendation score $y_{b}^u$ for the ACRecCos model is then computed by changing the right hand side of Equation~\ref{eq:acrec_v2} to read:
\begin{equation*}
\mathbf{w}_3 f(\mathbf{W}_2 f (\mathbf{W}_1[\mathbf{lp}^u_t; \mathbf{sp}_t^u; \mathbf{s}_b; \mathbf{ac};cos(\mathbf{ac}, \mathbf{d}_b)]))
\end{equation*}
Given that incorporating Cosine improved the performance of ACRec, in the rest of the paper we use ACRec to refer to the recommendation model that utilizes cosine as a feature.

\begin{table*}[htb]
    \centering 
    \small
    \begin{tabular}{cccccccccccc}
        \cline{3-12}
        & \multicolumn{11}{c}{Number $m$ of latest items for short-term history} \\
        \cline{3-12}
        & {} & 15 & 20 & 25 & 30 & 35 & 40 & 45 & 50 & 55 & 60  \\
        \hline
        \multirow{2}{*}{ACRec-2} & {HR@10} & 79.2 & 81.6 & 80.4 & {\bf 82.4} & 80.0 & 77.9 & 82.0 & 81.3 & 79.5 & 75.6 \\
                                & {NDCG@10} & 53.9 & 53.4 & 53.9 & {\bf 56.3} & 52.6 & 51.5 & 55.1 & 55.2 & 52.6 & 49.9 \\
        \hline
    \end{tabular}
    \caption{Performance of ACRec-2 on validation time steps without long-term preference.}
    \label{tab:ae-m} 
\end{table*}

\section{Baselines}
\label{sec:baselines}

For all baseline models, we follow the hyperparameter settings, training procedures, and fine-tuning strategies as described in their respective papers. A brief overview of each baseline is provided below.

\begin{itemize}
    \item {\bf SASRec} \cite{sasrec} is an ID-based sequential recommender model that applies unidirectional self-attention to capture user-item interaction patterns over time. Each item is represented by a learned ID embedding combined with a positional embedding, without relying on textual features.

    \item {\bf BERT4Rec} \cite{bert4rec} is an ID-based model that employs bidirectional self-attention with masked language modeling (MLM) to learn user and item embeddings without using textual content. It models user sequential behavior by randomly masking items and predicting them based on both left and right context.

    \item {\bf UniSRec} \cite{unisrec} is a text-based model that uses item descriptions instead of ID embeddings to learn item representations with a pre-trained language model like BERT. It applies bidirectional self-attention and contrastive pretraining to learn universal sequence representations, enabling transfer across domains using parametric whitening and a Mixture-of-Experts adaptor. 

    \item {\bf RecFormer} \cite{recformer} is a text-based model that formulates each item as a "sentence" by flattening key-value attribute pairs {\bf (e.g., title, category, brand, color)} into natural language input, entirely eliminating the use of item IDs. It uses a bidirectional Transformer based on Longformer to learn item and sequence representations from text, and is pre-trained with masked language modeling and item-item contrastive tasks.

    \item {\bf LLaRA} \cite{llara} is a hybrid model that integrates item ID-based embeddings from a traditional sequential recommender (e.g., SASRec) with item text metadata {\bf (e.g., title)} using a trainable projector to align both into the input space of a pretrained LLM. It uses {\bf LLaMA 2-7B} as the language model backbone and applies {\bf LoRA} for parameter-efficient fine-tuning. To effectively incorporate both textual and behavioral signals, LLaRA employs curriculum prompt tuning, which begins with text-only prompts and progressively introduces behavioral tokens during training. The following is the original movie recommender prompt used by LLaRA:

    \begin{itemize}
    \item LLaRA Movie Recommender Prompt: "This user has watched [HistoryHere] in the previous. Please predict the next movie this user will watch. Choose the answer from the following 20 movie titles: [CansHere]. Answer: "
    \end{itemize}
    
    For training and evaluation of the LLaRA model in our book recommendation setting, we adapted the prompt as follows:
    
    \begin{itemize}
        \item LLaRA Book Recommender Prompt: "This user has read [HistoryHere] in the previous. Please predict the next book this user will read. Choose the answer from the following 20 book titles: [CansHere]. Answer: "
    \end{itemize}

    \begin{itemize}
        \item {\bf LLaRA-AC:} To laverage users’ AC statements, we modified the prompt accordingly and trained the model using the same user–item interaction steps as in the ACRecCos model. During training, we utilized a combination of A, C, and AC phrases and we then evaluated the model on both the A+C+AC and the A+AC combination. The prompt we used here is as follows:

        \begin{itemize}
            \item LLaRA-AC Prompt: "This user has read [HistoryHere] in the previous. Now this user wants to read a book that has the following characteristics: [ac\_phrases]. Please predict the next book this user will read. The book title candidates are [CansHere]. Choose the answer from the following 20 book titles: [CansHere]. Answer: "
        \end{itemize}
    \end{itemize}

    \item {\bf iLoRA} \cite{ilora} is an extension of LLaRA that integrates a {\bf Mixture of Experts (MoE)} mechanism into the standard LoRA framework to address user behavior variability in sequential recommendation. Rather than applying a uniform LoRA module across all sequences, iLoRA dynamically assembles instance-specific LoRA modules by splitting the low-rank adaptation matrices into multiple experts. We train and evaluate iLoRA using the same procedure as LLaRA. The following is the default book recommendation prompt format used in iLoRA:
    
    \begin{itemize}
        \item iLoRA Book Recommender Prompt: "This user has read [HistoryHere] in the previous. Please predict the next book this user will read. The book title candidates are [CansHere]. Choose only one book from the candidates. The answer is "
    
    \end{itemize}
    To leverage AC statements, we extend this prompt format to form {\bf iLoRA-AC}, which includes the user’s desired affective-cognitive expectations:
    
    \begin{itemize}
        \item {\bf iLoRA-AC} Prompt: "This user has read [HistoryHere] in the previous. Now this user wants to read a book that has the following characteristics: [ac\_phrases]. Please predict the next book this user will read. The book title candidates are [CansHere]. Choose only one book from the candidates. The answer is "
    
    \end{itemize}
    
\end{itemize}

\section{Hyperparameter Tuning, Model Size and Computing Infrastructure}
\label{sec:tuning}

We implement our proposed model utilizing the Pytorch framework in Python.  We perform a grid search on validation examples for: dropout in $\{ 0.1, 0.2, 0.3, 0.4, 0.5\}$, learning rate in $\{0.1, 0.01, 0.001, 0.0001, 0.00001\}$, number of encoder blocks in $\{1, 2, 3, 4, 5, 6\}$, the number $m$ of items for modeling users' short-term preferences in $\{15, 20, 25, 30, 35, 40, 45, 50, 55, 60\}$. Upon performing the grid search, the dropout is set to 0.2, the learning rate to 0.0001, the number of encoder blocks to 4, and the number of latest items to 30 for our model. The hidden dimension in the transformer encoder block is 128. We also tried hidden dimensions of 64, 256, 512, however, the model with a dimension size of 128 gave the best results on validation data. Correspondingly, the total number of parameters in the ACRec model is 551,294. The hyperparameter tuning and experimental evaluations were conducted in a high-performance computing cluster on a A40 GPU with 48 GB memory, running for approximately 40 hours.

\section{Mapping Affective Statements to the Emotion Wheel}
\label{sec:wheel-classification-appendix}

To support the user Selection of AC statements described in Section~\ref{sec:acrec-task}, all statements that contain affective content (A + AC) were mapped to the 26 + 1 categories on the Emotion Wheel. For this task, we employed GPT-4o to assign each affective (A or AC) statement to one or more of 26 primary emotion categories. The prompt also instructed the model to consider not only a set of typical words for that emotion category, but also synonymous words or expressions that convey similar emotional meanings, to improve classification coverage. Any AC statement expressing a reader emotion not falling into the 26 predefined categories was labeled as "Other." If a phrase expressed multiple distinct emotions, it was assigned to all relevant categories within the taxonomy.

The full prompt used for this classification is available at the following link\footnote{\url{https://github.com/Ton-moy/ACRec_Taxonomy}}. To assess the quality of this emotion-based categorization, we manually annotated a random sample of 100 phrases and compared the results to GPT-4o's outputs. The model achieved an accuracy of 93\%. The distribution of AC statements with affective content across the 26 emotion categories is summarized in Table~\ref{tab:emotion_distribution}.

\begin{table}[ht]
\centering\small
\begin{tabular}{lr}
\toprule
Emotion Category &  Count \\
\midrule
        Interest &            3310 \\
         Sadness &            1341 \\
        Surprise &            1256 \\
    Feeling love &            1027 \\
       Amusement &            1012 \\
      Wonderment &             935 \\
        Pleasure &             910 \\
         Longing &             800 \\
      Compassion &             744 \\
             Joy &             643 \\
            Hope &             620 \\
           Other &             591 \\
         Tension &             579 \\
            Fear &             557 \\
     Contentment &             312 \\
  Disappointment &             165 \\
           Anger &             155 \\
          Relief &             120 \\
      Relaxation &             107 \\
         Disgust &              93 \\
       Gratitude &              71 \\
          Hatred &              53 \\
           Pride &              35 \\
           Guilt &              21 \\
            Envy &              17 \\
        Contempt &              17 \\
           Shame &              16 \\
\bottomrule
\end{tabular}
\caption{Emotion Category Distribution.}
\label{tab:emotion_distribution}
\end{table}


\end{document}